\documentclass[12pt]{article}
\usepackage{tgtermes}
\usepackage{geometry}
\usepackage[utf8]{inputenc}
\usepackage{xcolor}
\usepackage{soul,color}

\usepackage[%
  giveninits=true,
  style=chem-angew,
  backend=bibtex,
  doi=false,
  isbn=false,
  url=false,
  citestyle=numeric, sorting=none,maxbibnames=3]{biblatex}
\AtEveryBibitem{%
  \clearfield{pages}%
}
\usepackage[flushmargin,para]{footmisc}

\defbibenvironment{bibliography}
  {\noindent}
  {\unspace}
  {\printtext[labelnumberwidth]{%
    \href{\thefield{url}}{\bf \printfield{labelnumber}}}\addspace}{}
    
\renewbibmacro*{finentry}{\finentry\addspace}

\usepackage{enumitem,amssymb}
\usepackage{ragged2e}
\usepackage{pifont}
\usepackage{sectsty}
\usepackage{titlesec}

\usepackage[colorlinks=true]{hyperref}
\hypersetup{
urlcolor=blue,
citecolor=black,
linkcolor=black
    }

\sectionfont{\large}
\subsectionfont{\normalsize}
\titlespacing\section{0pt}{6pt}{4pt}
\titlespacing\paragraph{0pt}{4pt}{2pt}
\begin{document}

\huge
\begin{center}
Critical Laboratory Studies to Advance Planetary Science and Support Missions
\linebreak
\linebreak
\large
A white paper submitted to the Planetary Science and Astrobiology \linebreak Decadal Survey 2023-2032
\linebreak
\end{center}
\normalsize

\begin{flushleft}

Edith C. Fayolle$^{1 \star}$, Laurie Barge$^1$, Morgan Cable$^1$, Brian Drouin$^1$, Jason P. Dworkin$^2$, Jennifer Hanley$^3$, Bryana L. Henderson$^1$, Baptiste Journaux$^4$, Aaron Noell$^1$, Farid Salama$^5$, Ella Sciamma-O'Brien$^5$, Sarah E. Waller$^1$, Jessica Weber$^1$, Christopher J. Bennett$^6$ J\"urgen Blum$^7$, Murthy Gudipati$^1$, Stefanie Milam$^2$, Mohit Melwani-Daswani$^1$, Michel Nuevo$^5$, Silvia Protopapa$^8$, Rachel L. Smith$^9$\\[\baselineskip]

$^1$ NASA Jet Propulsion Laboratory, California Institute of Technology\\
$^2$ NASA Goddard Space Flight Center\\
$^3$ Lowell Observatory \& Northern Arizona University\\
$^4$ University of Washington\\
$^5$ NASA Ames Research Center\\
$^6$ University of Central Florida\\
$^7$ Inst. f\"ur Geophysik und extraterrestrische Physik, Techn. Univ. Braunschweig\\
$^8$ Southwest Research Institute\\
$^9$ North Carolina Museum of Natural Sciences; Appalachian State University\\
$^{\star }$ Contacts: \href{mailto:edith.c.fayolle@jpl.nasa.gov}{\nolinkurl{edith.c.fayolle@jpl.nasa.gov}}, +1-626-487-7392\\[\baselineskip]

The list of {\bf co-signatories} can be found here: \href{https://docs.google.com/spreadsheets/d/12ch2XWRUCJNIxHTmF7rl1QHR8SvQg7Acxbt4lDL1YeY/edit?usp=sharing}{\nolinkurl{https://docs.google.com/spreadsheets/d/12ch2XWRUCJNIxHTmF7rl1QHR8SvQg7Acxbt4lDL1YeY/edit?usp=sharing}}\\[\baselineskip]

This paper was initiated and is endorsed by members of the \href{https://lad.aas.org/}{Laboratory Astrophysics Division} of the American Astronomical Society.\\[\baselineskip]

\vspace*{\fill}

A part of this work was carried out at the Jet Propulsion Laboratory, California Institute of Technology, under a contract with the National Aeronautics and Space Administration (80NM0018D0004). \copyright 2020, All rights reserved.
\end{flushleft}

\thispagestyle{empty}  
\pagebreak
\setcounter{page}{1}

\noindent Laboratory studies for planetary science and astrobiology aim at advancing our understanding of the Solar System through the promotion of theoretical and experimental research into the underlying processes that shape it. Laboratory studies (experimental and theoretical) are crucial to interpret observations and mission data, and are key incubators for new mission concepts as well as instrument development and calibration. They also play a vital role in determining habitability of Solar System bodies, enhancing our understanding of the origin of life, and in the search for signs of life beyond Earth, all critical elements of astrobiology. Here we present an overview of the planetary science areas where laboratory studies are critically needed, in particular in the next decade. These areas include planetary \& satellites atmospheres, surfaces, and interiors, primitive bodies such as asteroids, meteorites, comets, and trans-Neptunian objects, and signs of life. Generating targeted experimental and theoretical laboratory data that are relevant for a better understanding of the physical, chemical, and biological processes occurring in these environments is crucial. For each area we present i) a brief overview of the state-of-the-art laboratory work, ii) the challenges to analyze and interpret data sets from missions and ground-based observations and to support mission and concept development, and iii) recommendations for high priority laboratory studies.

\paragraph{Overall recommendations} The paper advocates that:
\vspace{-4pt}
\begin{itemize}
\setlength\itemsep{-4pt}
\item Increased and steady support for fundamental laboratory studies for planetary science (both experimental and theoretical) and astrobiology among the various agencies is critical. 
\item Explicit support and inclusion of laboratory studies by missions and projects are essential to enable mission formulation and the analysis of returned data to maximize scientific output.
\item Strong Instrumentation, Technology, and Facilities Development Programs are needed to support the development, construction, and maintenance of state-of-the-art laboratory instrumentation and facilities.
\item Adequate funding for critically evaluated new and existing databases which are needed for the analysis of observations, {\it in situ} measurements, and the modeling of planetary environments.

\end{itemize}
\vspace{-4pt}
We also wish to stress the importance of maintaining a vibrant community of scientists conducting laboratory studies through faculty development to ensure the health and vitality of the field.

\section{Planets and satellites: atmospheres}
\paragraph{Composition}
From radio to ultraviolet wavelengths through infrared and visible, spectrometers and radio receivers on ground-based and remote space observatories enable the detection of spectral signatures of atoms and molecules in (exo)planet and satellite atmospheres and provide information on their composition. Physical characteristics of atmospheres such as winds and global circulation can also be gleaned from high fidelity measurements of gas-phase species. Experimental and theoretical studies can provide the spectra of atomic and molecular species (neutral and ionic) in relevant environments for direct comparison to ground-based and spacecraft observations. Exploring the effects on these spectra of the wide ranges of physical parameters (pressure, temperature, etc.) occurring in planetary environments, represents a fundamental cornerstone of planetary science \cite{tinetti2013}. Space missions can also probe planetary atmospheres {\it in situ} with mass spectrometry (MS), which helps with the identification of specific neutral and ionic molecular species. Novel, quantitative laboratory spectroscopy and MS techniques that push to state-of-the art sensitivity and/or calibration methodologies continue to be needed to directly compare to observations and feed into increasingly accurate quantum and statistical mechanical frameworks. Theoretical efforts are also needed to complement the experimental efforts, by providing consistent advancement and continual improvement of extrapolations to remote environments beyond experimental capability. {\bf These laboratory and theoretical data then need to be parameterized consistently and provided, in standard formats via databases (e.g. hitran.org, astrochemistry.org/pahdb, webbook.nist.gov) for general use in order to interpret observations and use as inputs for models. Long-term funding to support databases is needed.}

\paragraph{Reactivity}
Planetary atmospheres are exposed to various energy sources (solar UV radiation, electrons, ions, galactic cosmic rays, etc.), that can induce dissociation and ionization of the main molecular constituents, and result in the formation of larger molecular compounds, solid aerosols, and cloud particles. Combined critical laboratory, modeling, and theoretical efforts have been conducted for several decades to simulate the atmospheric chemistry in planetary atmospheres \cite[e.g.][]{horst2017}. This work has paved the way for future investigations of atmospheric chemistry in Solar System bodies and exoplanets, and has demonstrated the importance of interdisciplinary studies to advance our understanding. For example, in the case of Titan, the largest moon of Saturn, this combined approach has revealed the importance of the ionic species in the gas phase chemistry leading to the production of aerosols in the atmosphere \cite{vuitton2019}, hence {\bf demonstrating the need for future ionic and neutral MS measurements with higher resolution and wider mass range (observations and experiments). Along with these MS data, measurements of new and improved chemical reaction rates involving positive and negative ions and larger molecules, at varying temperatures and pressures representative of different environments are also needed to improve atmospheric models.}. The importance of complex molecules such as aromatic species in atmospheric chemistry has also been unveiled by synergistic studies of chemical pathways linking aromatics to the formation of aerosols \cite[][and references therein]{horst2017}, demonstrating that more work on aromatic spectroscopy in the gas and solid phase is needed to better assess their impact on the chemistry processes at work.

\paragraph{Aerosols}

Atmospheric aerosols play a key role in determining the thermal structure and climate. They also serve as sinks for gas-phase species and can act as condensation nuclei for the formation of clouds. They can also interact with the surface materials and contribute to the surface composition. {\bf There is a critical need for chemical and physical characterization of analogs of atmospheric aerosols and cloud particles (experiments and theory) relevant to various (exo)planetary atmospheric compositions (Titan, Pluto, Venus, giant planets, exoplanets) to better model their interaction in planetary atmospheres and on surfaces in order to predict their impact on observations.} In particular, optical constants from the UV to the Far IR of analogs of atmospheric aerosols and cloud particles produced in different experimental conditions relevant to various (exo)planetary environments where refractory materials have been detected are critically needed. They are further used as input parameters in radiative transfer models for global circulation simulations and to produce synthetic spectra to be compared to observations \footnote{see white papers (WPs) by Roser et al. and Kohler et al.}. Additionally, {\bf theoretical modeling efforts are also critically needed to accurately predict aerosol formation and chemistry} especially in environmental conditions difficult to simulate in a laboratory setting (different pressures, temperatures, timescales).

\section{Planets and satellites: surfaces}
\paragraph{Characterization}
Laboratory work is imperative to understand surface composition of other planetary bodies. Without reference spectra collected in a controlled environment, identification of ices, minerals, and organics would be incredibly challenging. Laboratory studies must match the conditions of the planetary surface (e.g. temperature, pressure, and atmospheric composition) in order to give the best results. To date, spectral libraries have primarily consisted of Earth-relevant materials measured at Earth-relevant temperatures. Studies have shown the importance of measuring materials at the correct conditions \cite[e.g.][]{hanley2014}, as spectra will show vastly different features at different temperatures or with different grain sizes. Laboratory studies are also needed to understand the stability of materials under various conditions (temperature, pressure, radiation, etc). For instance, under the cold, dry Martian conditions, salts might become dehydrated, so it is important to study all the phases (including hydration states) of these minerals and ices. Another challenge in surface composition characterization is the issue of mixing materials. {\bf Spectra of mixed materials cannot be represented by simple linear combinations of their components and characterization of mixed materials is needed.} Surface composition models also need spectral data at the proper temperatures and with control of grain size, mineral hydration level, and layering (i.e., due to precipitation or sedimentation). Laboratory studies are just beginning to tackle this massive experimental landscape \cite[e.g.][]{protopapa2015,lynch2015}. Laboratory studies support spectroscopic identification across multiple platforms, from field work to landed missions, and from orbiters to Earth-based telescopes. Reference spectra must be of high enough spectral resolution across broad wavelength regions to satisfy all these observations. Each laboratory typically focuses on a specific wavelength region since one can learn different, though complementary, information from them. But the quality of spectral fitting be improved by collecting (and fitting) over a wide spectral range. {\bf Spectral characterization of different materials and mixtures under well-controlled experimental conditions is crucial to understand these materials and building a robust, comprehensive spectral library.}

\paragraph{Processing}

Surfaces are affected by changes in temperature, particle weathering (photons, electrons, protons and heavier ions), micrometeorite impact gardening, and interactions (physical and chemical) with the atmosphere and interior. Understanding how this processing occurs – and quantifying it – is necessary to comprehend the evolution of surfaces, which, when combined with interior modeling, could provide insight into hard-to-reach planetary interiors, and help with mission design and development of {\it in situ} detection instruments. The interpretation of laboratory studies on surface processing is challenging since the interaction of even a few components can quickly become complex. Also, processes occurring on geological timescales can only be studied on a shortened timescale, which can limit direct laboratory to planetary surfaces comparison. To overcome this difficulty and to complement these studies, a mechanistic and quantitative understanding of the underlying physical or chemical processes is needed. Experiments involving fewer components are conducted under various conditions to derive fundamental parameters (rates of thermal and radiation-induced reactions, sputtering rates and products, desorption and diffusion processes, etc.) which are then used as inputs in simulations to understand more complex systems. {\bf In light of the many lander missions and concepts, it is important to predict the chemical complexity that could be encountered on surfaces via formation \& destruction rate measurements and simulations.} This will help set thresholds for instrument sensitivity and sample collection depth requirements\footnote{see WP by Choukroun et al.}. An understanding of surface mechanical properties under analogous temperature, radiation, and gravitational conditions is also crucially needed to help predict landing conditions and develop appropriate drilling and sampling strategies. The exploration of ocean worlds would also benefit from a deeper understanding of liquid-solid chemistry at cryogenic temperatures, salt formation and weathering. Finally, the development of lunar bases and the need for resource extraction would greatly benefit from fundamental sublimation-distillation work.

\section{Planets and satellites: interiors}

Processes happening in interiors of planetary bodies affect observable surface features, atmosphere, magnetosphere, or lack thereof. Our understanding of these processes solely relies on indirect observations (e.g. gravity, magnetic, seismological measurements) and inverse modeling, based on sometimes sparse laboratory data. Thus, increasing our knowledge on material properties at interior conditions will greatly reduce uncertainties in interior inverse modeling. This will also directly increase the scientific return from interior science driven missions and instruments.

\paragraph{Composition}

Thermodynamics of aqueous systems (water mixed with volatiles and salts) and their interactions with silicates at high pressure are arguably the most important fields of study to better understand the interiors and the habitability of icy satellites in the outer Solar System. Progress in high-pressure experimental technologies (e.g. diamond anvil cells, hydraulic pressure vessels, laser shocks) and {\it in situ} analytical techniques (e.g. Raman, sound speed, X-ray diffraction) allows the reproduction of these extreme environments and the measurement of properties such as density, composition, equilibria, and thermal properties. The last ten years of high-pressure experimental studies on icy worlds have shown the vast lack of accurate data at the relevant conditions, even on simple systems such as water and its numerous high-pressure ice polymorphs \cite{journaux2020}. Furthermore, the unusual pressures and temperatures found inside these bodies are known to allow for the formation of new phases with unique properties (e.g. clathrates and salt hydrates). A myriad of phases, potentially major components of planetary interiors, remain to be discovered and characterized. Therefore, {\bf experimental studies of relevant chemical systems (aqueous, organics, silicates, and their mixtures) at extreme conditions are needed and would allow for a better characterization of petrological equilibria, thermodynamic properties (e.g. density, heat capacity, thermal expansivity) and transport properties (e.g. electrical conductivity, heat conductivity).}

\paragraph{Thermal-mechanical properties and dynamics}

Surface activity features, detected by remote sensing, are investigated by inferring the dynamic and geological evolution of satellites, which are constrained by geodynamic models based on the known mechanical properties of materials at the relevant conditions. Although silicate rheology has been studied for decades for Earth science applications, the mechanical properties of ices, hydrates and their assemblages remain poorly investigated experimentally. The effect of high pressures in large ocean-worlds cannot be correctly extrapolated from data collected in well-studied ranges below 100~MPa. As an example, very few studies have characterized and quantified the creep behavior of high-pressure phases and assemblages, including ice polymorphs, at the small strain rates occurring in icy ocean worlds. Furthermore, the effects of grain size, temperature, impurities and recrystallization processes are currently unconstrained by experimental data. {\bf Therefore, new laboratory studies of slow strain rate creep behavior of ice polymorphs, clathrates, salt hydrates and their assemblages at low and high pressure covering the range of conditions found in the icy satellites of the Solar System are crucial.} Due to the long-term nature of these experiments (from weeks to several months per run), adequate length of support will be imperative.
\paragraph{Oceans and interaction with the cores}

Laboratory experiments have been used to mimic chemical processes occurring in oceans, at ocean-seafloor, and ocean-ice interfaces, to inform habitability models. Chemical production of new species is investigated under various conditions including mineral type simulating the seafloor, temperature, pH, redox conditions, as well as the presence of iron and other catalytic metals. Perhaps the most difficult condition to reproduce in the laboratory is extremely high pressure. Only a few research groups worldwide are equipped with reactors capable of safely investigating liquid chemistry under relevant high pressure and temperature conditions, and simulate chemistry near hydrothermal vents. Since the pressure greatly affects the chemical and mineral interactions and properties, new experimental reactors using gold bags or titanium are being designed \cite[e.g.][]{shibuya2015}. It is additionally important for these experiments to have proper techniques to analyze the materials produced - isotopic labelling, for example, can assist with the assignment of organics and the mechanisms of their production. Reaction rates and products are precious inputs for numerical thermodynamic simulations that predict and quantify the occurrence of chemical reactions and habitability in the interior of ocean worlds, small bodies, and Early Mars \cite[e.g.][]{canovas2020}. The last ten years have seen a large increase in the number of mission concepts that target ocean worlds and aim at looking for life that could initially develop in their sub-surface oceans. {\bf Support for the development of high-pressure liquid simulation chambers is needed to inform chemical models of sub-surface oceans and assess their habitability potential.}

\paragraph{Plume characterization}
Active jets resulting in plumes at Enceladus, and very likely also at Europa, present a unique opportunity to sample sub-surface ocean/reservoir composition without the need for a lander/melt probe to get through the often kilometers-thick ice crust. The Ion and Neutral Mass Spectrometer and Cosmic Dust Analyzer instruments onboard the Cassini spacecraft sampled Enceladus' plume. These instruments were, however, not designed to quantify biomolecules and other indicators of life following hypervelocity impact. Laboratory studies have since aimed at replicating the observed mass spectral patterns to infer the plume's composition. Techniques include cryogenic/metal projectiles fired at surfaces or laser dispersion experiments. The latter accurately reproduces ice grain impact mass spectra and identifies and quantifies salts and organics in the plume grains \cite{postberg2018}. These techniques, however, do not accurately reproduce the impacts of individual and ensemble ice grains as they would occur on a mission. New experiments are being developed to generate, accelerate, and impact ice grains (both single grains and ensembles as well as charged and neutral species) in order to interrogate their post-impact mass spectra with respect to velocity. Theoretical models have also been generated to predict how the energy of impact is redistributed to volatilizing, ionizing, and fragmenting ice grain constituents, with particular focus on the survivability of organics ensconced within the grain \cite{jaramillo2012}. With the Surface Dust Analyzer instrument on Europa Clipper and the mission concepts developed for Enceladus and Triton, {\bf investing in experimental facilities and theoretical frameworks that can simulate hypervelocity impacts of neutral species and generate mass spectral libraries is crucial.} They would also enable flight instrument testing and validation for these essential life-searching missions.

\section{Primitive bodies}
\paragraph{Asteroids and meteorites} Meteorites, interplanetary dust particles, and asteroids contain evidence of the chemical and geological conditions encountered during the Solar System formation and on some evolved bodies \cite{abreu2018}. The recent detection of organics on the surface of Ceres during the Dawn mission and on Themis-family bodies has provided a glimpse into the rich asteroids chemistry. Laboratory and modeling studies of interior and surface chemistry on these bodies are needed to both interpret current observations and to support future {\it in situ} missions \footnote{see WP by Castillo-Rogez et al., Jacobson et al.}. The collection of meteorites for scientific analysis is the most economical method of planetary materials collection and tens of thousands of samples have been collected so far\footnote{\href{https://www.lpi.usra.edu/meteor/}{lpi.usra.edu/meteor}}. {\bf It is vital that atmospheric dust collection, micrometeorite collection, and field expeditions—particularly the Antarctic Search for Meteorites—continue}. Likewise, it is vital that these samples continue to be curated, catalogued, and distributed to the community. They are a bridge to the interpretation of asteroid sample return material (e.g. Hayabusa, OSIRIS-REx) and {\it in situ} analyses (e.g. Psyche, Lucy), and sample return material puts the meteorite and dust collection in context. Spurred by organic sample return missions, the importance of extraterrestrial organics, and the unique challenges of organic and biological contamination, {\bf investments in controlling and understanding contamination in meteorite collections must continue. Investments in R\&A laboratories for astromaterial analysis to maintain state-of-the-art equipment is necessary.} Long-term steady investments are needed in both multiuser facilities and individual principal investigator laboratories for equipment, method development, and staffing for the analysis of both sample return material as well as meteorites and dust \cite{NAS_sample_2019}. Spectral databases (e.g. RELAB) are important for comparing meteorite spectra with Solar System objects. However, the samples are often rare or otherwise have not been studied with modern instrumentation. An effort to crosslink databases of spectral features and detailed chemical composition should be made.

\paragraph{Comets}

The composition and morphology of comets is of great interest due to the processing they encounter during their journey in the Solar System and their potential at unveiling the chemical makeup of the pre-Solar Nebula. The chemical composition of the comet 67P/C-G sublimating nucleus reveals a striking similarity with that seen around sun-like protostars \cite{drozdovskaya2019}. Remote sensing and {\it in situ} measurements from the ROSETTA mission rely on spectroscopic and MS laboratory standards for species detections \cite[e.g.][]{mumma2011,goesmann2015}. While spectroscopic, chemical, and surface sublimation studies are still needed to understand comets’ chemical make-up from observations and their pre-solar origins, the next big step in comet exploration will likely involve a cryogenic sample return. {\bf Laboratory experiments under realistic conditions (low gravity, high porosity, low temperature, high-vacuum) and accompanying improved modeling of the nucleus are needed} to understand mechanical properties such as compaction and sintering effects, as well as chemical and physical processes {\bf to inform sample extraction techniques and the estimated depth needed to reach pristine material.} Challenges associated with sample alteration during the return phase as well as curation and analysis also heavily rely on R\&A laboratories and are detailed elsewhere\footnote{see WP by Milam et al., Westphal et al.}.

\paragraph{Trans Neptunian Objects (TNOs)} 

Since these objects are less affected by sublimation than comets, they offer a good opportunity to probe the Solar Nebula conditions, despite their observation and exploration being more difficult due to their distance. The New Horizons mission had a tremendous impact on our understanding of TNOs and highlighted the role that small TNOs can play in solving the Solar System formation puzzle \cite{mckinnon2020}. Future JWST observations and concepts to Triton and further into the Solar System will likely bring us a wealth of information on the composition of TNOs. {\bf Composition identification relies, however, on experiments that measure optical constants of icy surfaces.} Spectra of some of the likely ice components have already been measured, especially in the mid-IR for molecular physics or astrophysical purposes, but {\bf data are still missing in the near-IR and UV to inform TNOs observations}. Combined modeling and quantitative experimental efforts on the pre-Solar Nebula chemistry\footnote{see WP by \href{https://113qx216in8z1kdeyi404hgf-wpengine.netdna-ssl.com/wp-content/uploads/2019/05/518_gudipati.pdf}{Gudipati et al.} in Astro2020 for laboratory studies relevant to star- and planet-forming regions} and on the weathering of TNOs’ surfaces also have the potential to date surfaces and identify initial formation regions in the Nebula, thus providing clues on the Solar System assembly. Finally, the shape of a couple of TNOs is also planned to be observed by JWST during occultation events. {\bf Simulations and laboratory experiments on the sticking of ice and dust grains of different sizes \cite[e.g.][]{Gundlach2014}, are thus needed to inform models of planetesimal coagulation and planetary bodies formation.}

\section{Signs of life}

Laboratory studies play a key role in guiding astrobiology-focused missions on the best places and techniques to look for signs of life, as well as how to interpret the signatures identified. This work should focus on three key areas: {\bf (1) Understanding the abiotic formation and background levels of potential biomolecular targets; (2) Constraining the limits of survival, growth, and adaptation of Earth microbes in controlled laboratory analog environments; and (3) Interpreting the preservation and detection of biosignatures at locations accessible to missions.} Continued measurements of rate constants and reaction pathways leading to the formation and degradation of compounds of biological relevance (amino acids, nucleobases, etc.) and their initial polymerization mechanisms are needed to improve the ability of models to predict what abiotic background scenarios may look like. This could more precisely demonstrate the deviations from the abiotic scenario which would indicate the presence of biology (cf WP by Barge et al.). Studies should also continue to explore how organisms adapt to survive and thrive under extreme conditions (e.g. low temperatures, high pressure, radiation, briny environments, etc.), working under controlled conditions to identify the limits of that adaptability, with organisms and experiments that can lead to important mechanistic insights into biological response to different stressors. Lastly, significant efforts are required to quantify how conditions at mission-accessible locations impact whole organisms and the biosignatures that they are built from. Studies are needed to understand whether biosignatures can maintain their structures, distributions, and characteristic biological imprints and how these can be reliably detected and interpreted. It is essential to evaluate whether these biochemical systems can be well-characterized with the sampling and measurement techniques developed for {\it in situ} habitability missions, and to ensure that these techniques are reliable and robust in the given environments. Collectively, all three of these laboratory thrusts will allow a significant enhancement to Astrobiology mission planning and science return for the next decade.

\section*{References}

\printbibliography[heading=none]

\end{document}